\documentclass[12pt]{iopart}
\usepackage{iopams}  
\usepackage{graphicx}
\usepackage{color}
\usepackage{multicol}
\usepackage{float}
\usepackage{wrapfig}

\begin{document}
\title[Laser-ion acceleration by nano-layered flat-top cone targets]{Modeling the interaction of an ultra-high intensity laser pulse with nano-layered flat-top cone targets for ion acceleration}
\vspace{2pc}
\author{O Budrig\u{a}$^1$, E d'Humi\`{e}res$^2$, L E Ionel$^1$, M Marti\c{s}$^1$ and M Caraba\c{s}$^3$}

\address{$^1$ Laser Department, National Institute for Laser, Plasma and Radiation Physics, M\u{a}gurele, Romania}
\address{$^2$ Universit\'{e} de Bordeaux - CNRS - CEA, CELIA, Talence, France}
\address{$^3$ Faculty of Automatic Control and Computers, University POLITEHNICA of Bucharest, Bucharest, Romania}
\eads{\mailto{maria.boghet@inflpr.ro},\mailto{olimpia.budriga@inflpr.ro}}

\begin{abstract}
We propose new nanotargets as nano-layered flat-top cone targets to be used for future laser-ion acceleration experiments at ELI-NP. We study their behaviour at the interaction with an ultra-high intensity laser pulse by performing Particle-In-Cell simulations. We analyze spatio-temporal the electromagnetic field based on the finite-difference time-domain method for a complementary description. We find the optimum diameter of the nanospheres and the proper nano-flat-top foil thickness for which one can obtain monoenergetic beams of very energetic ions with low angular divergence.  
\end{abstract}

\pacs{52.38.Kd}
\vspace{2pc}
\noindent{\it Keywords\/}: LASER-ION ACCELERATION, ULTRA-HIGH INTENSITY LASER PULSE, NANO-LAYERS FLAT-TOP CONE TARGET, RADIATION PRESSURE ACCELERATION, PLASMA

\maketitle

\normalsize
\section{Introduction}
At Extreme Light Infrastructure Nuclear Physics (ELI-NP) are envisioned experiments with the two ten Petawatts power lasers for laser-ion acceleration with applications in medicine such as hadron therapy \cite{Eliwhitebook,Tanaka}. In the last decade a lot of target geometries were proposed in order to obtain very energetic protons which can be used to distroy tumors  \cite{Bulanov2002,Esirkepov2002,Fourkal2002,Malka2004,Tajima2009,Bulanov2015}. Some works showed that the interaction of the ultra-high intensity laser pulse with a micro-cone target can generate protons accelerated at energies of tens MeV with low angular divergence and high laser absorption \cite{Flippo2008,Nathalie2010,Gaillard2011,Olimpia2017} in the Target Normal Sheath Acceleration and Direct Laser-Light-Pressure regimes. Other papers are devoted to different kinds of cone targets suitable for proton acceleration at energies up to few tens of MeV \cite{Sentoku2004,Zhou2010,Yu2012,Wu2013,Bake2016,Yang2017}. A new experiment from Vulcan facility shows that the interaction of the ultra-high intensity laser pulse with a ultra-thin foil can generate protons accelerated at energies exceeding 94 MeV \cite{Higginson2018}.    

We propose two new types of nanotargets for laser-ion acceleration. First type of nanotargets is a flat-top cone target \cite{Flippo2008,Olimpia2017} with a nano-layer foil in the tip of the cone. The second type is a flat-top cone target with a flat-top foil consisting of two nano-layers. First nano-layer is composed by nanospheres with the same diameter and tangential to each other and the second one is a nano-foil. We perform two-dimensional (2D) Particle-in-Cell (PIC) simulations for the interaction of the ultra-high intensity laser pulse with the plastic targets described above in order to find the optimum nanosphere diameter and nano-flat-top foil thickness to obtain the highest ion energies, the most monoenergetic ion beam and the smallest angular divergence of the ion beam. The laser pulse has the similar parameters with the two high-power lasers from ELI-NP \cite{whitebookELINP}. Also, a spatio-temporal analysis of the electromagnetic field distribution during the interaction of the high-intensity laser pulses with the nano-layered flat-top cone targets has been performed using a commercial software (RSoft, by Synopsys Optical Solution Group). This study is based on finite-difference time-domain (FDTD) method for solving Maxwell equations and it was used to determine the electromagnetic field in the vicinity of the femtosecond pulses$-$nano-layered flat-top cone target interaction point for different nano-flat-top foils and nanospheres dimensions. A lot of FDTD studies had been elaborated in order to determine the optimum conditions for extreme electromagnetic field generation \cite{Laura2016,Laura2014,Lin2017}. This method aims to provide a complementary approach to the PIC simulations.

We find that for the ultra-high intensity laser pulse circularly polarized the protons have the maximum energy higher than 1 GeV. The localized accelerated protons beam is monoenergetic with the protons energy greater than 200 MeV for all investigated nano-flat-top cone targets. Also, the localized carbon C$^{6+}$ ions beam has the energy at least 4 GeV. Therefore, we achieve that the mono and double nano-layer flat-top cone targets can be used in the hadron therapy. In the Section \ref{laser_target} of the paper we describe the laser-pulse and the target characteristics. The parameters of the PIC simulations and the results of them are presented in Section \ref{PIC}. We analyse the proton and carbon ion acceleration in Sections \ref{prot_acc} and \ref{ion_acc}, respectively. The numerical simulations based on FDTD method and their results are shown in Section \ref{elemagFDTD} and they are compared with the PIC simulations results. We conclude our work with some remarks in the Section \ref{concl}. 

\normalsize

\section{Laser-pulse and target description}
\label{laser_target}
\begin{figure}
\vspace{5mm}
\centering
 \includegraphics[width=4cm]{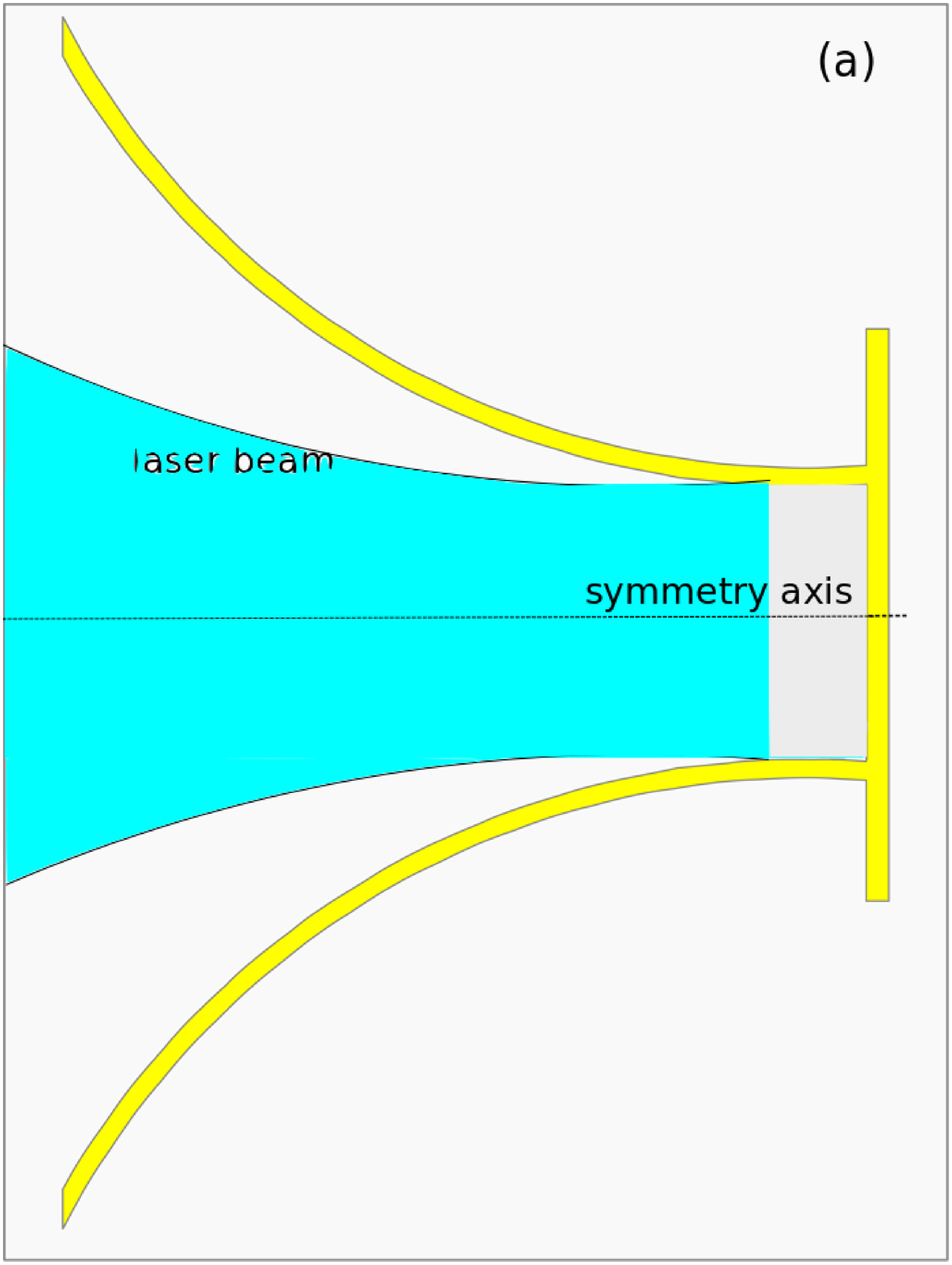}\label{fig1a}
\hspace{5mm}
 \includegraphics[width=4cm]{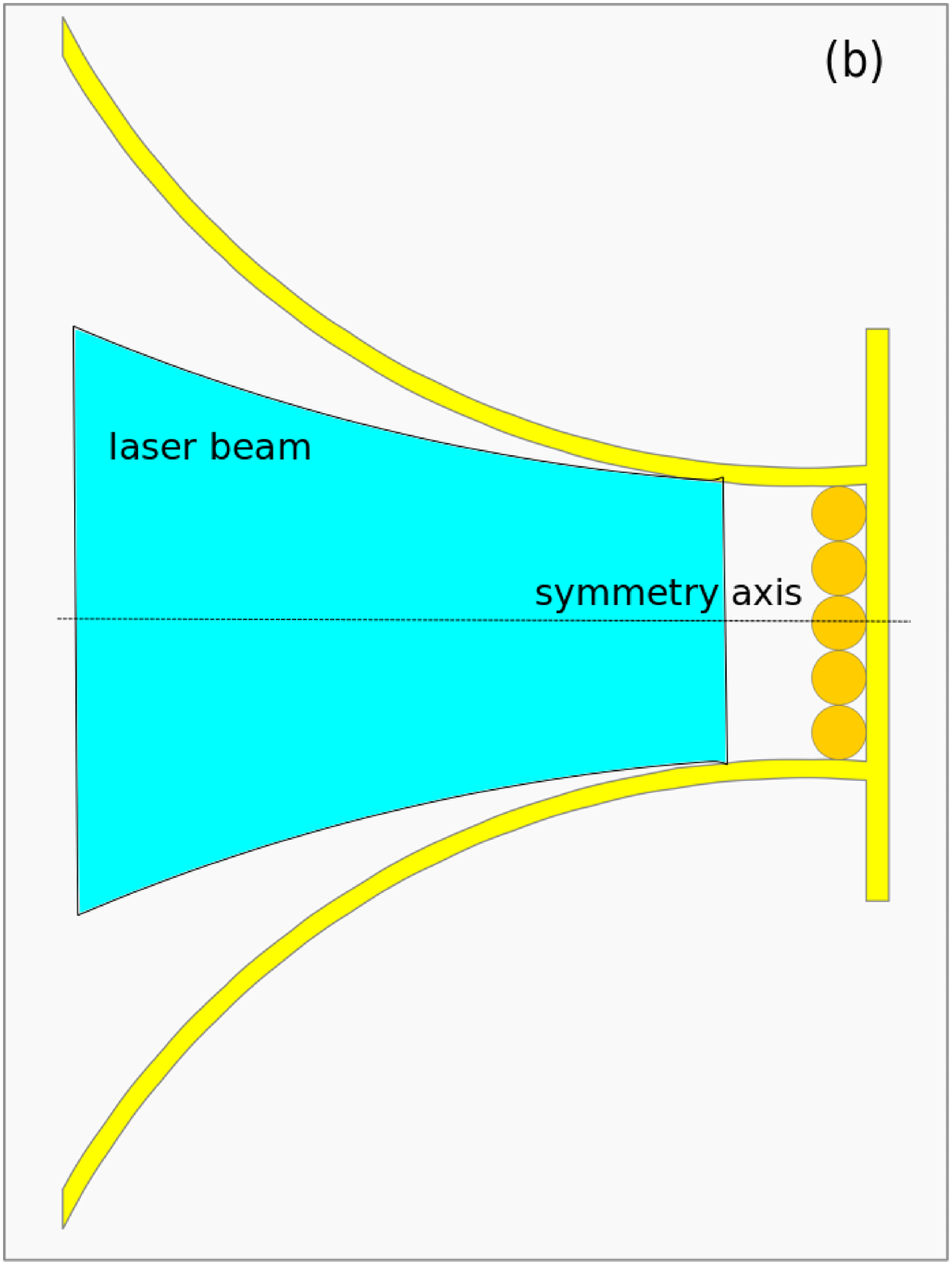}\label{fig1b}
\caption{The geometry of (a) the nano-layer flat-top cone target and (b) nanospheres layer on a nano-layer flat-top cone target.}
\label{fig:1}      
\end{figure}

As we mentioned above, we used in our Particle-in-Cell simulations an ultra-high intensity laser-pulse with parameters similar to  those of the two 10 PW lasers from ELI-NP \cite{whitebookELINP}. In all simulations from this work the ultra-high intensity laser pulse has a Gaussian profile with a central wavelength $\lambda_{0}=800$ nm, the period $\tau_{0}$ = 2.7 fs, the duration $\tau$ = 25 fs, a laser focal spot FWHM $d = 5.6 \mu$m and an intensity peak $I=2.16\times 10^{22}$ W/cm$^2$. Therefore, the normalized laser pulse amplitude is a$_{0}=100$, the laser pulse energy and power are E = 532 J and P = 21.3 PW, respectively. The ultra-high intensity laser pulse propagates along the symmetry axis of the flat-top cone target as can be seen in Figures \ref{fig:1}(a) and \ref{fig:1}(b). Hence, the incidence angle of the laser beam on the flat-top foil from the tip of the flat-top cone target is zero degree. The $x$-axis is chosen parallel with the symmetry axis of the flat-top cone target. We consider both linearly $p$-polarized ($E_{z}=0$ and $E_{y}\neq 0$) and circularly polarized ($E_{z}=E_{y}\neq 0$) laser pulse, where $E_{z}$ and $E_{y}$ are the $z$-component and $y$-component, respectively of the electric field.

We drew the geometry of the target which consists in a plastic nano-layer flat-top cone target (see Figure \ref{fig:1}(a)) and nanospheres layer on a nano-layer flat-top cone target (Figure \ref{fig:1}(b)). All nanospheres have the same diameter and are tangential one to other. The laser pulse alaways comes from the left. The cone height is 20 $\mu$m, the base diameter is 25 $\mu$m, the cone neck diameter is 5 $\mu$m, the walls thickness is 4 $\mu$m and the width of the flat-top foil is 15 $\mu$m. These cone targets can be printed in our days with commercial printers by two-photon polymerization process on nano-foils and nano-layered foils. The nano-foils and nano-layered foils can be nano-films.          

\section{PIC simulations}
\label{PIC}
We intend to achieve the optimum thickness of the flat-top nano-layer foil and the nanospheres diameter such that to have protons accelerated at hundreds of MeV in monoenergetic beams with very low angular divergence. For this propose we used two dimensional version of the 3D relativistic PIC code PICLS \cite{Sentoku2008}. The PIC simulations have been performed on the cluster from the High Performance Computing Center, Faculty for Automatic Control and Computers, University POLITEHNICA of Bucharest, Romania \cite{Mihai2017}.

We choose a simulation box with the dimensions of 60 $\mu$m x 75 $\mu$m (6000 cells x 7500 cells). The time of simulation is 330 fs. The grid step-size of the simulations is 10 nm and the time step is 0.033 fs.
The plasma has the density $n_{e}=320n_{c}$, where the critical plasma density is $n_{c} = 1.72\cdot 10^{21}$ cm$^{-3}$. The plasma is totally ionized, i.e. with carbon ions C$^{6+}$ due to ultra-high laser pulse intensity of $2.16\times 10^{22}$ W/cm$^2$. Therefore, we put in each grid cell 21 electrons, 3 protons, 3 C$^{6+}$ ions. 

\subsection{Proton and carbon ion acceleration}
\label{prot_acc}
Within this work we intend to obtain very monoenergetic and very collimated accelerated protons beam for the treatment of the deep inside tumors. In order to distroy tumors the protons from the monoenergetic beam must have at least 200 MeV kinetic energy \cite{Bulanov2002,Esirkepov2002,Fourkal2002}. 

\begin{figure}
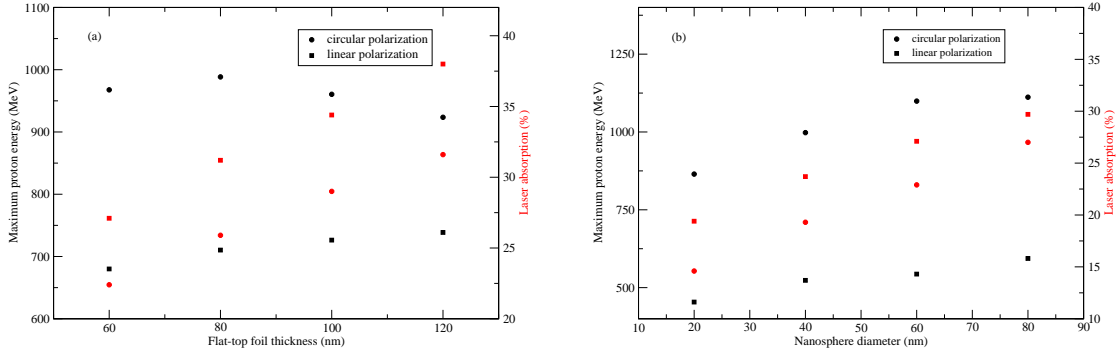

\vspace{5mm}
\centering
 \includegraphics[width=7cm]{Fig2a.eps}\label{fig2a}
\hspace{5mm}
 \includegraphics[width=7cm]{Fig2b.eps}\label{fig2b}
\caption{Maximum proton kinetic energy (black points) and laser absorption (red points) as a function of (a) the nano-foil thickness for a plastic nano-layer flat-top cone target and (b) the nanospheres diameter for a plastic double nano-layered flat-top cone target.}
\label{fig:2}      
\end{figure}

First we perform 2D PIC simulations to study the interaction of the ultra-high intensity laser pulse with a flat-top cone target which have in the tip a plastic nano-layer foil. The thickness of the nano-layer foil has the values of 60, 80, 100 and 120 nm. We choose the thickness of the nano-layer flat-top foil (nano-film) such as to be able to compare it with the below case of the nanospheres layer on a nano-layer foil flat-top cone target. The maximum proton energy versus the nano-layer foil thickness is represented with black points in the Figure \ref{fig:2}(a) for linear and circular polarization of the ultra-high intensity laser pulse. The highest value of 988 MeV for the maximum proton energy is obtained in the case of the circularly polarized (CP) ultra-high intensity laser pulse for the thickness of the nano-layer flat-top foil equal to 80 nm. In the case of the linearly polarized (LP) ultra-high intensity laser pulse which interacts with a 120 nm flat-top foil thickness of a nano-layer flat-top cone target we have the highest maximum energy of the protons equal to 739 MeV. We can see that the maximum proton energy for the circularly polarized laser pulse (circles) is higher than the one for the linearly polarized laser pulse (squares) at any value of the nano-layer foil thickness. The laser absorption on the particle acceleration shown by the red points in Figure \ref{fig:2}(a) has a linear dependence on the flat-top foil thickness. Laser absorption represents the rate of the laser pulse energy which is transferred in the kinetic energy of the protons, ions and electrons and is calculated as the ratio between the maximum total kinetic energy and the initial laser pulse energy \cite{Olimpia2017}. The laser absorption is greater for the linearly polarized than for the circularly polarized ultra-high intensity pulse.           

For the flat-top cone with a nano-foil coated with nanospheres we varied the nanospheres diameter while we keep constant the thickness of the nano-foil equal to 40 nm. In Figure \ref{fig:2}(b) we plot with black points the dependence of the maximum proton kinetic energy on the nanospheres diameter for both linearly and circularly polarized ultra-high intensity laser pulse. We consider the diameter of the nanospheres having the values of 20, 40, 60 and 80 nm. We can see that the maximum proton energy has the values higher for a laser pulse with circular polarization (circles) than the linear polarization (squares). The dependence of the maximum proton kinetic energy on the nanospheres diameter is not linear. In the Figure \ref{fig:2}(b) we depict with red color the laser absorption. It can be seen that for the circularly polarized ultra-high intensity pulse the laser absorption is lower than for the linearly polarized laser pulse.   

\begin{figure*}
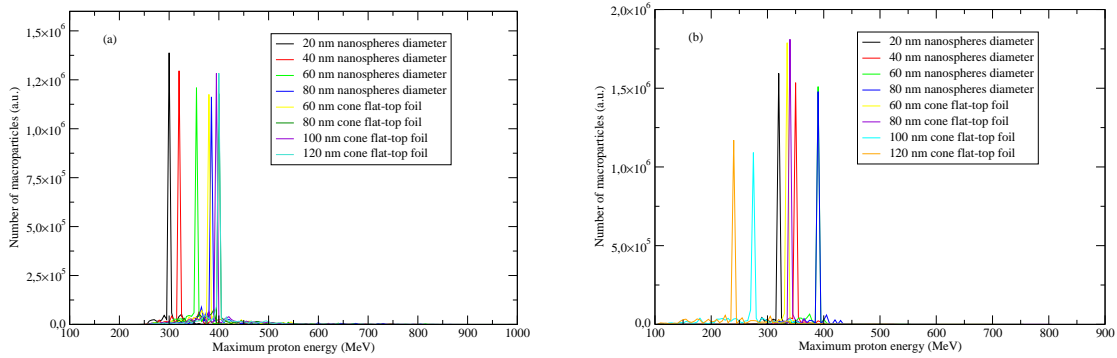

\vspace{5mm}
\centering
 \includegraphics[width=7cm]{Fig3a.eps}
\hspace{5mm}
 \includegraphics[width=7cm]{Fig3b.eps}
\caption{Proton energy spectra for a plastic nano-layered flat-top cone target and an ultra-high intensity laser pulse (a) circularly and (b) linearly polarized at the time of simulation 67$\tau_{0}$. The protons have kinetic energy greater than 100 MeV and are localized between 40 and 56 microns on $x$-axis.}
\label{fig:3}      
\end{figure*}

The localized protons spectra for all the investigated targets are shown in Figure \ref{fig:3}. From Figures \ref{fig:3}(a) and \ref{fig:3}(b) we can see that the number of protons accelerated at a kinetic energy higher than 100 MeV and located between 40 and 56 microns on $x$-axis are a little higher for linear polarization than for circular polarization of the ultra-high intensity laser-pulse. 
\begin{multicols}{2}
\end{multicols}
The beams of the energetic protons are monoenergetic and have the kinetic energy centered between 200 and 400 MeV for both laser pulse polarizations. The nano-layered flat-top cone target having a nano-layer foil with the thickness equal to 80 and 120 nm produced a monoenergetic beam of protons with the highest energy of 400 MeV at the interaction with a CP ultra-high intensity laser pulse (see Figure \ref{fig:3}(a)). It is shown in Figure \ref{fig:3}(b) that in the case of the LP of the ultra-high intensity laser pulse the most energetic proton beam with the energy of 390 MeV is obtained for the 80 nm nanospheres diameter layer on a 40 nm nano-foil flat-top cone target. For this flat-top cone target with 80 nm nanospheres layer and CP ultra-high intensity laser pulse we obtained the most energetic beam of all nanospheres diameter.       
    
\begin{figure*}
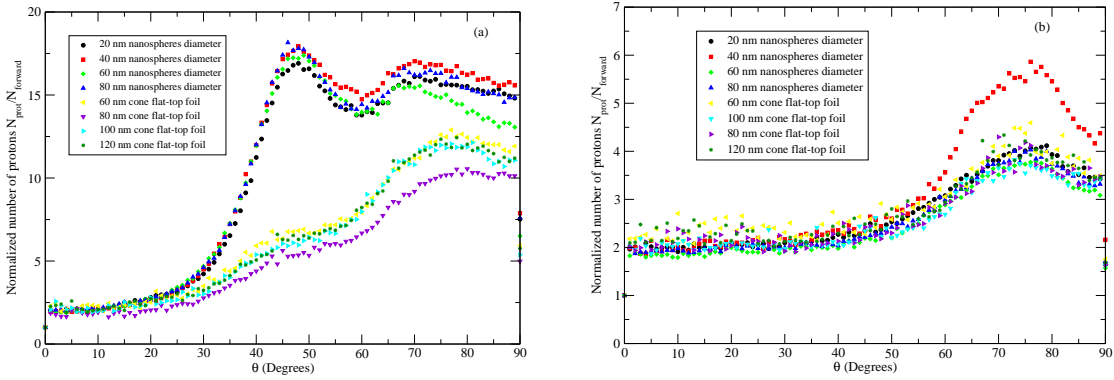

\vspace{5mm}
\centering
 \includegraphics[width=7cm]{Fig4a.eps}
\hspace{5mm}
 \includegraphics[width=7cm]{Fig4b.eps}
\caption{Normalized number of protons N\textsubscript{prot}/N\textsubscript{forward} as a function of proton divergence angle $\theta$ for a plastic nano-layered flat-top cone target interacting with a ultra-high intensity laser pulse (a) circularly and (b) linearly polarized at the time of simulation $t=67\tau_{0}$. The protons have kinetic energy between 100 and 500 MeV and are localized between 40 and 56 microns on $x$-axis.}
\label{fig:4}      
\end{figure*}

We calculated the ratio between the number of the protons accelerated under the $\theta$ angle, N\textsubscript{prot} and the number of the protons accelerated in the forward direction, N\textsubscript{forward}, to express how divergent is the proton beam. The proton divergence angle is defined $\theta = \arctan{(p_{y}/p_{x})}$, where $p_{x}$ and $p_{y}$ are the components of the proton momentum on the $x$ and $y$-axis, respectively. In Figure \ref{fig:4} is represented the normalized number of protons versus proton divergence angle for all investigated targets, protons with energies between 100 and 500 MeV and located between 40 and 56 $\mu$m at the time of simulation $67\tau_{0}$.    
\begin{multicols}{2}
\end{multicols}
The proton beams are less divergent for the flat-top cone targets with the flat-top foil composed by a nanospheres layer on a nano-foil than the flat-top cone targets with a nano-layer foil having the same thickness when interact with a CP ultra-high intensity laser pulse as can be seen in Figure \ref{fig:4}(a). For the nanosphere diameter of 80 nm the proton beam has the highest value of the normalized number of protons at the lowest value of the divergence angle of 46 degrees. The localized proton beams for all targets and a LP ultra-high intensity laser pulse are very divergent as it is shown in Figure \ref{fig:4}(b). The most of the protons are accelerated under an angle higher than 60 degrees. We must note that from our simulations we obtained a broad total protons energy spectra but very low divergent, with the divergence angle $\theta$ around few degrees for all analized targets.       

\subsection{Carbon ion acceleration}
\label{ion_acc}
\begin{figure*}
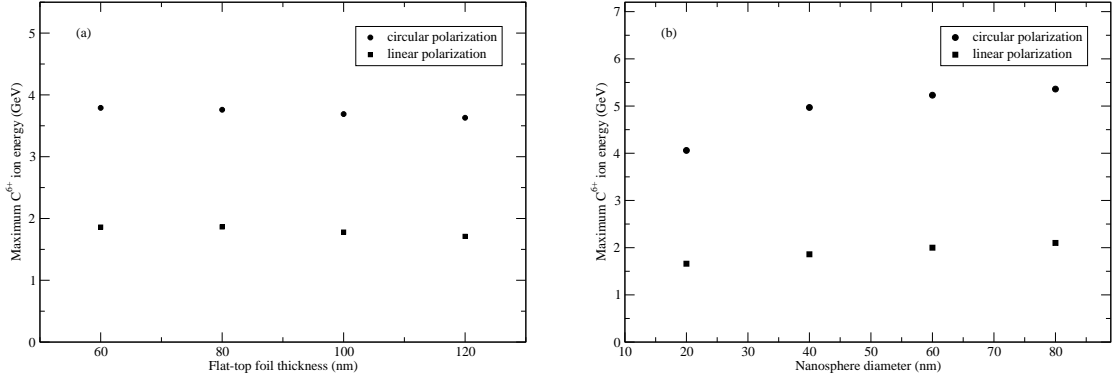

\vspace{5mm}
\centering
 \includegraphics[width=7cm]{Fig5a.eps}
\hspace{5mm}
 \includegraphics[width=7cm]{Fig5b.eps} 
\caption{Maximum kinetic energy of the carbon C$^{6+}$ ion as a function of (a) the nano-foil thickness for a nano-layer flat-top cone target and (b) the nanospheres diameter for a double nano-layered flat-top cone target.}
\label{fig:5}      
\end{figure*}

The carbon C$^{6+}$ ions accelerated in Break-Out Afterburner regime at 700 MeV (60 MeV/amu) \cite{JungNewJourPhys} and at 1 GeV (83 MeV/amu) \cite{JungPhysPlas} was evidenced experimentally for nm-scale targets and relativistic intensity laser pulse of 5 x 10$^{20}$ W/cm$^{2}$. Our simulated plasma contains carbon C$^{6+}$ ions, too. Therefore, we investigate how are accelerated the carbon C$^{6+}$ ions. For this purpose we calculate the maximum C$^{6+}$ ion kinetic energy and we plot it as a function of the nano-foil thickness and the nanospheres diameter in Figures \ref{fig:5}(a) and \ref{fig:5}(b), respectively. The ultra-high intensity laser pulse has linear and circular polarization.
\begin{multicols}{2}
\end{multicols}
We can see from the Figures \ref{fig:5}(a) and (b) that the maximum kinetic energies of the C$^{6+}$ ions are higher for the CP ultra-high intensity laser pulse than the LP ultra-high intensity laser pulse. This behaviour is the same with that of the maximum proton kinetic energy. The maximum kinetic energy of the carbon C$^{6+}$ ions is higher than 4 GeV in the case of a CP ultra-high laser pulse in interaction with a double nano-layered flat-top cone target. It is known that for hadron therapy of the tumors carbon C$^{6+}$ ions of 4-5 GeV in almost monoenergetic beams are nedeed \cite{Tajima2009}. 

\begin{figure*}
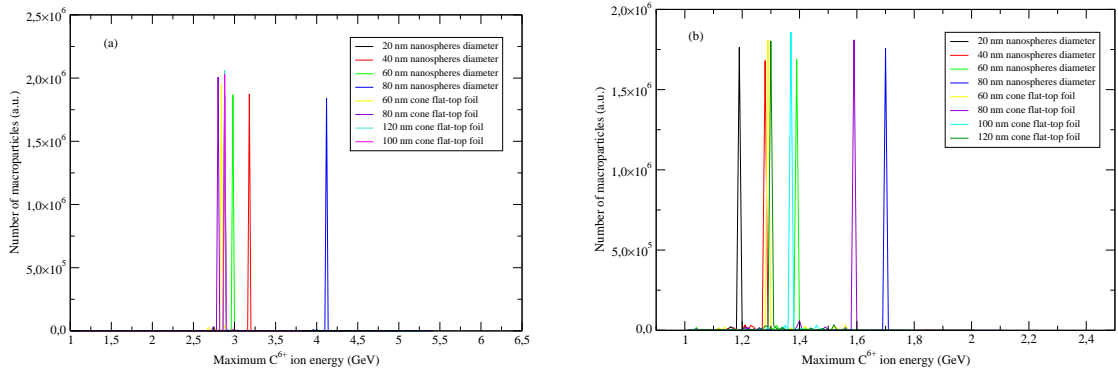

\vspace{5mm}
\centering
 \includegraphics[width=7cm]{Fig6a.eps}
\hspace{5mm}
 \includegraphics[width=7cm]{Fig6b.eps}
\caption{Carbon C$^{6+}$ ion energy spectra for (a) circular polarization and (b) linear polarization of the ultra-high intensity laser pulse at the time of simulation $t=67\tau_{0}$. Carbon C$^{6+}$ ions are localized between 40 and 56 $\mu$m on the $x$-axis.}
\label{fig:6}      
\end{figure*}

In Figures \ref{fig:6}(a) and \ref{fig:6}(b) are plotted the carbon C$^{6+}$ ion energy spectra for CP and LP ultra-high intensity laser pulse, respectively. We account only the carbon C$^{6+}$ ions with energies higher than 1 GeV, located on the $x$-axis between 40 and 56 $\mu$m at the time of simulation 67$\tau_{0}$ . 
\begin{multicols}{2}
\end{multicols}
All the carbon C$^{6+}$ ion energy spectra from the Figures \ref{fig:6}(a) and \ref{fig:6}(b) are monoenergetic. Only in the case of the CP ultra-high intensity pulse and the plastic double nano-layered flat-top cone target with a flat-top foil composed by a layer of 80 nm diameter nanospheres on a 40 nm foil the carbon C$^{6+}$ ion beam is monoenergetic with the carbon C$^{6+}$ ion  energy of 4.12 GeV (343 MeV/amu) (see Figure \ref{fig:6}(a)). Therefore, this double nano-layered flat-top cone target interacting with the CP ultra-high intensity pulse can be efficient in the hadron treatment of the malign tumors.  

\section{Electromagnetic field}
\label{elemagFDTD}
We studied the evolution of the electromagnetic field by both PIC and based on FDTD method simulations.

\begin{figure*}
\vspace{5mm}
\centering
\includegraphics[width=7cm]{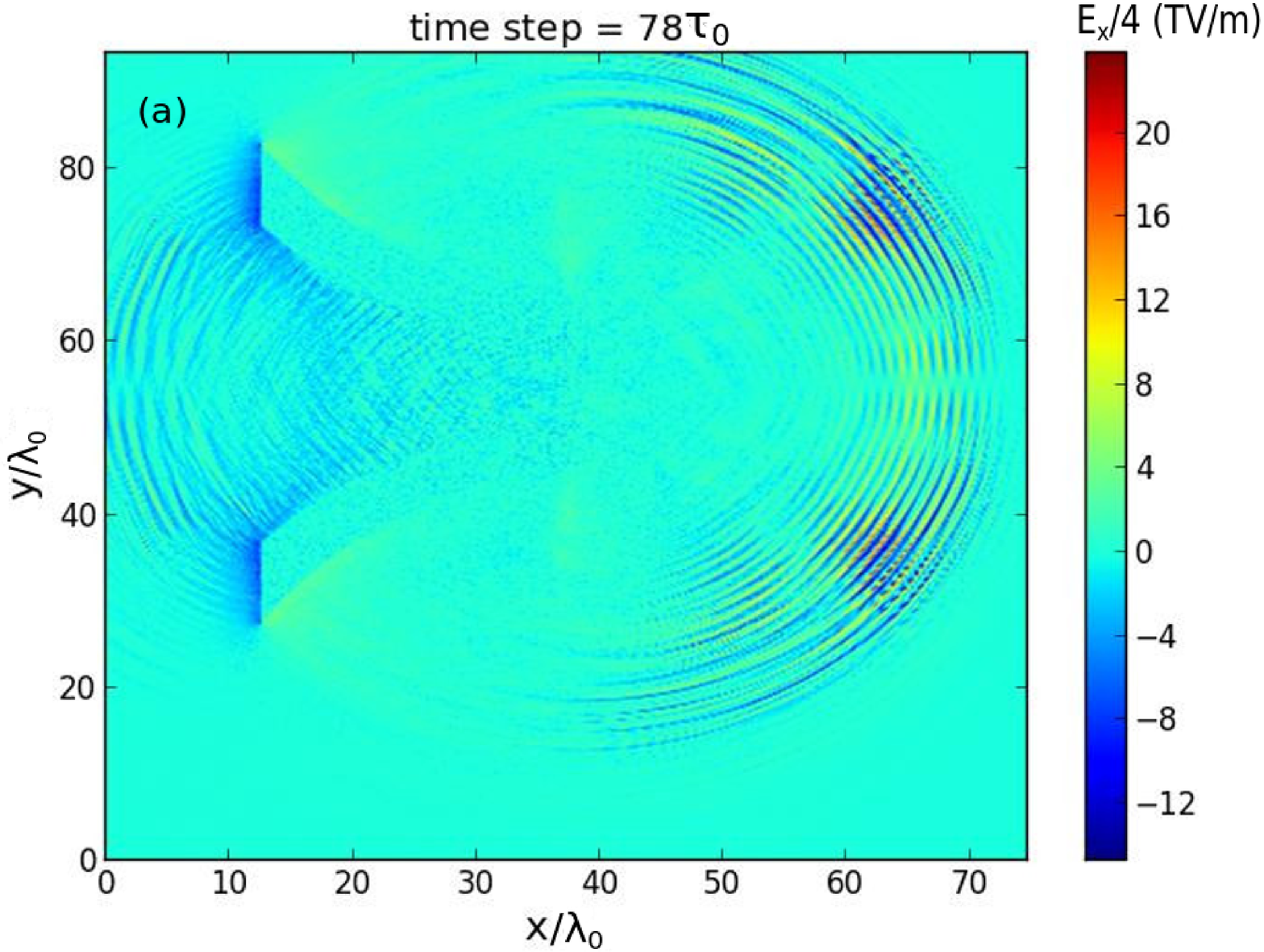}
\hspace{5mm}
\includegraphics[width=7cm]{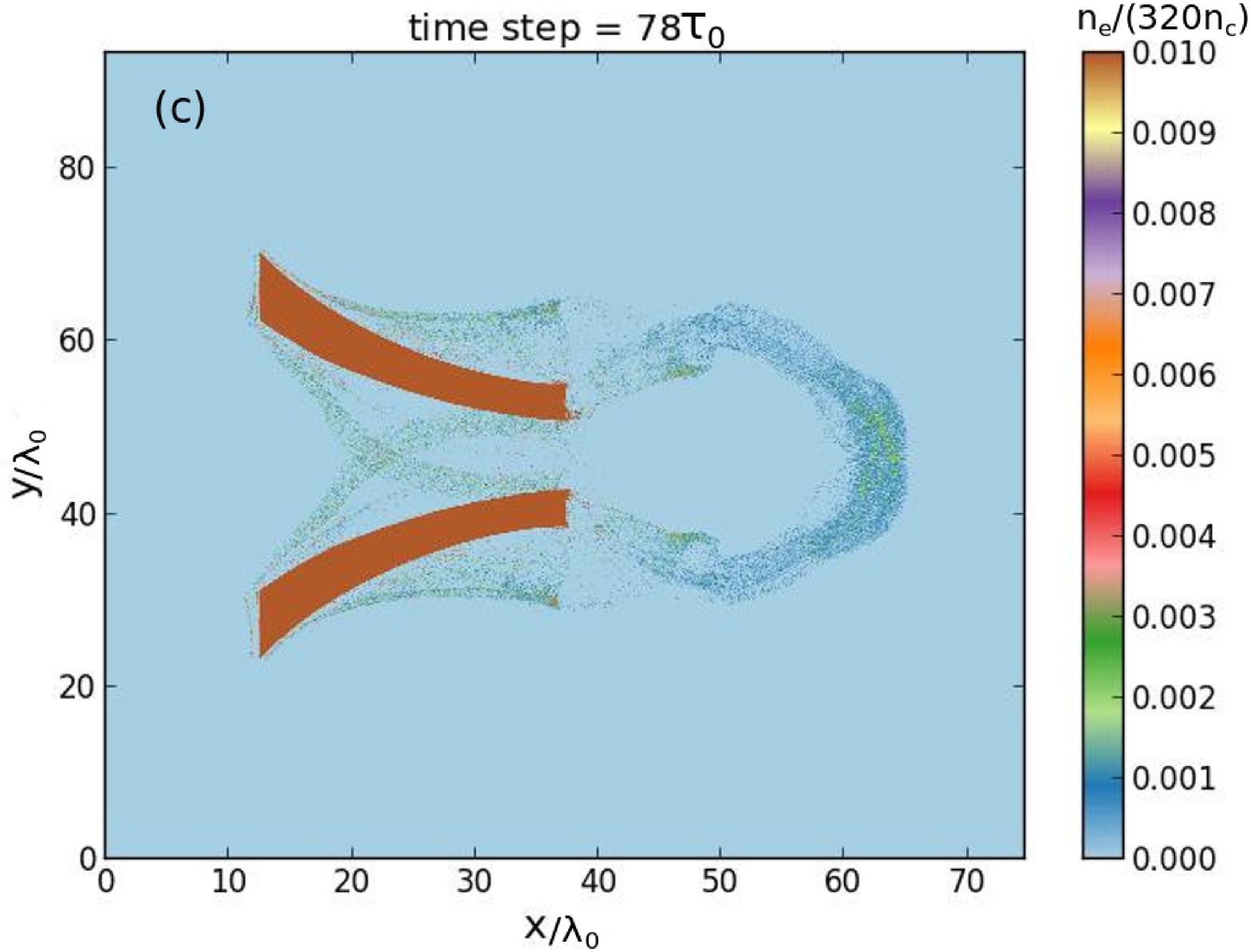} \\
\vspace{3mm}
\includegraphics[width=7cm]{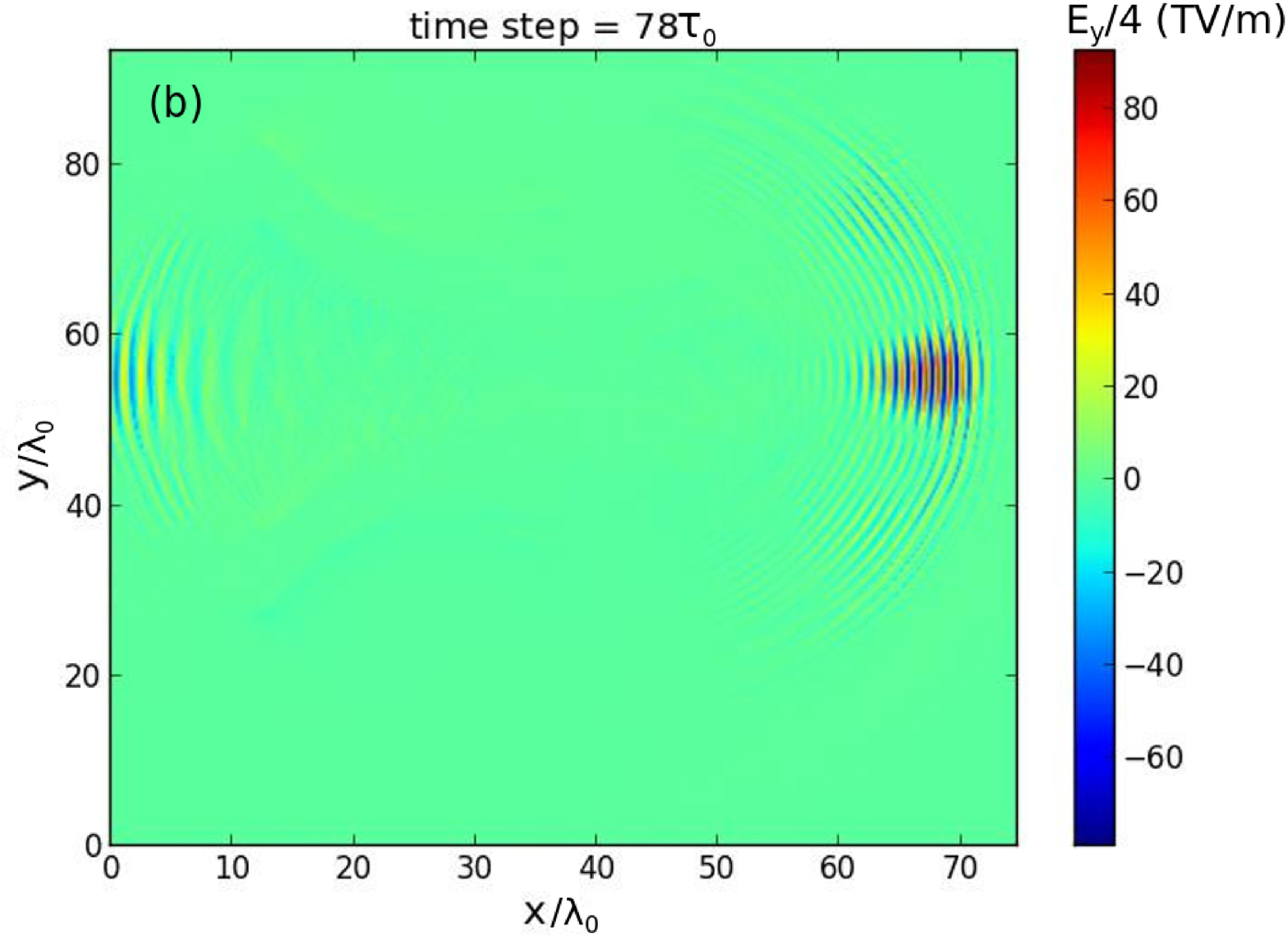}
\hspace{5mm}
 \includegraphics[width=7cm]{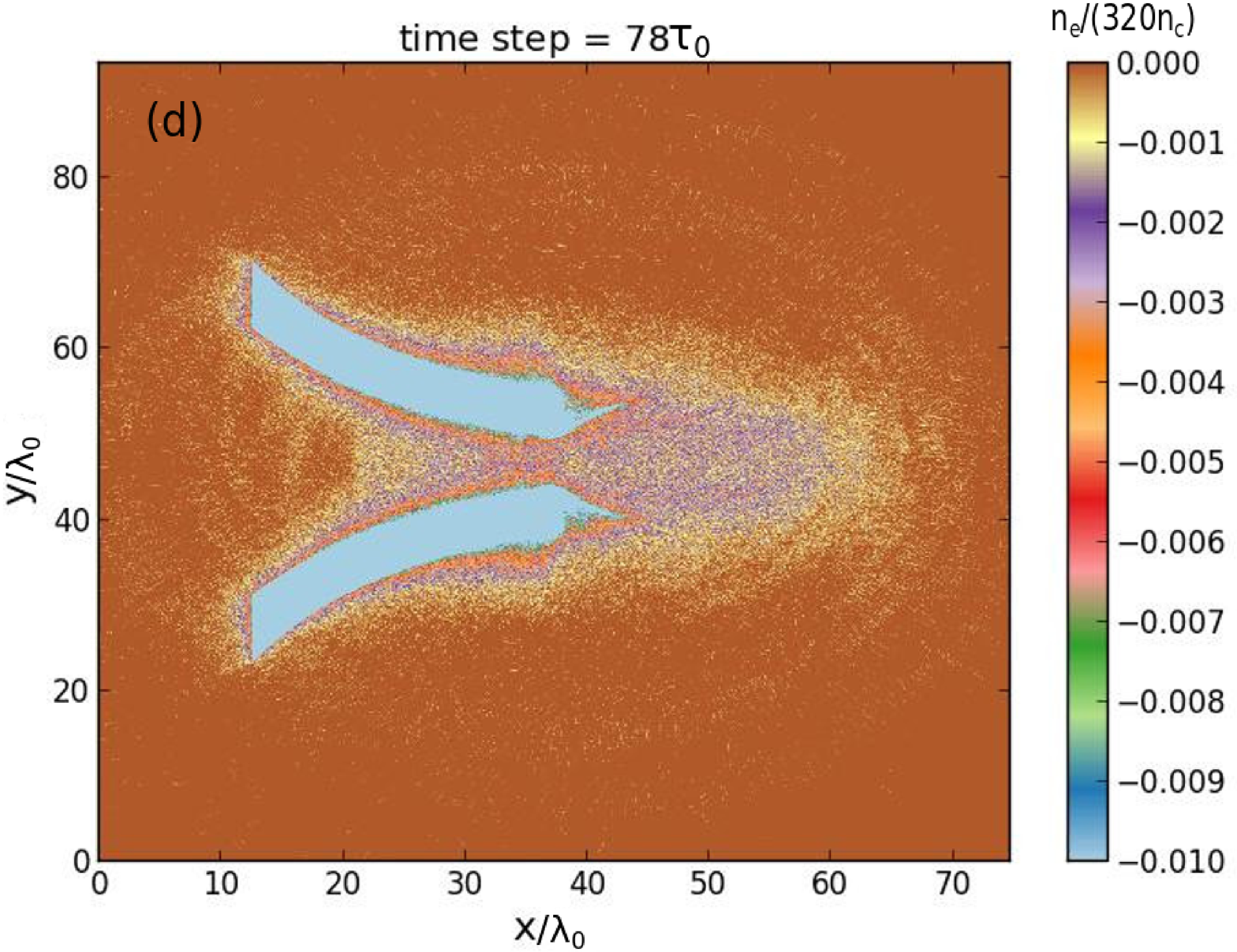} 
\caption{The electromagnetic field and the plastic flat-top cone target with the flat-top foil composed by a 80 nm diameter nanospheres layer on a nanofoil with a thickness of 40 nm at the time of simulation $t=78\tau_{0}$  plotted in {\it (x,y)} plane by (a) the $x$-component of the electric field, (b) the $y$-component of the electric field, (c) the proton density, and (d) the electron density.}
\label{fig:7}      
\end{figure*}

The electromagnetic field, proton density and electron density gives us significant informations about the regime of the laser ion acceleration regime as was highlighted in the previously papers \cite{Esirkepov2004,Silva2004,BulanovPlas2010,Zou2015}. From our 2D PIC simulations we can see that at $t=78\tau_{0}$ the ultra-high intensity laser pulse deforms the target in a cocoon which traps the electromagnetic field [Figures \ref{fig:7}(a)-\ref{fig:7}(d)]. It can be seen from Figure \ref{fig:7}(c) that in the front of the cocoon there is a dense sheet of protons which will be accelerated in the later time but will not evolves transversally as much as an ultra-thin foil target in the radiation pressure acceleration (RPA) regime \cite{BulanovPlas2010,Bulanov2010}. The electromgnetic field is reflected more and more in time by the plasma layer.  
\begin{multicols}{2}
\end{multicols}
This behaviour of the electromagnetic field, target evolution and the monoenergetic proton beams leads us to the conclusion that the laser-ion acceleration is in RPA regime \cite{Esirkepov2004,BulanovPlas2010,Zou2015,Bulanov2010,Chen2009,Yu2010}.

Numerical analysis of the the electric field distribution during the interaction of femtosecond laser pulses with nano-layered flat-top cone targets has been performed by using a commercial software (RSoft, by Synopsys Optical Solution Group). The numerical model solves the Maxwell equations based on the finite difference time domain (FDTD) method and it was used to determine the field amplitude in the vicinity of the femtosecond pulses $-$ cone targets interaction point for different nano-layer foil and nanospheres dimensions.

\begin{figure}
\vspace{5mm}
\centering
 \includegraphics[width=8cm]{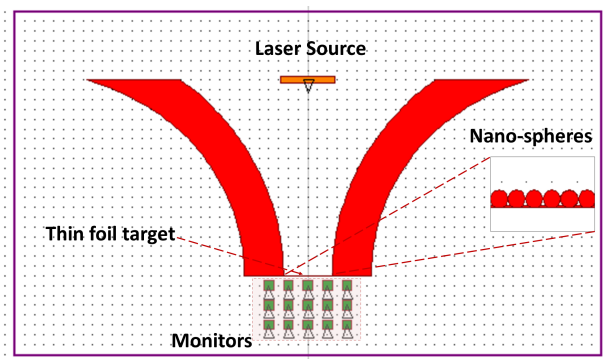}
\caption{Sketch of the ultra-high intensity laser pulse  flat-top cone target interaction geometry in FDTD simulations.}
\label{fig:8}      
\end{figure}

\begin{figure*}
\vspace{2mm}
\centering
 \includegraphics[width=7cm]{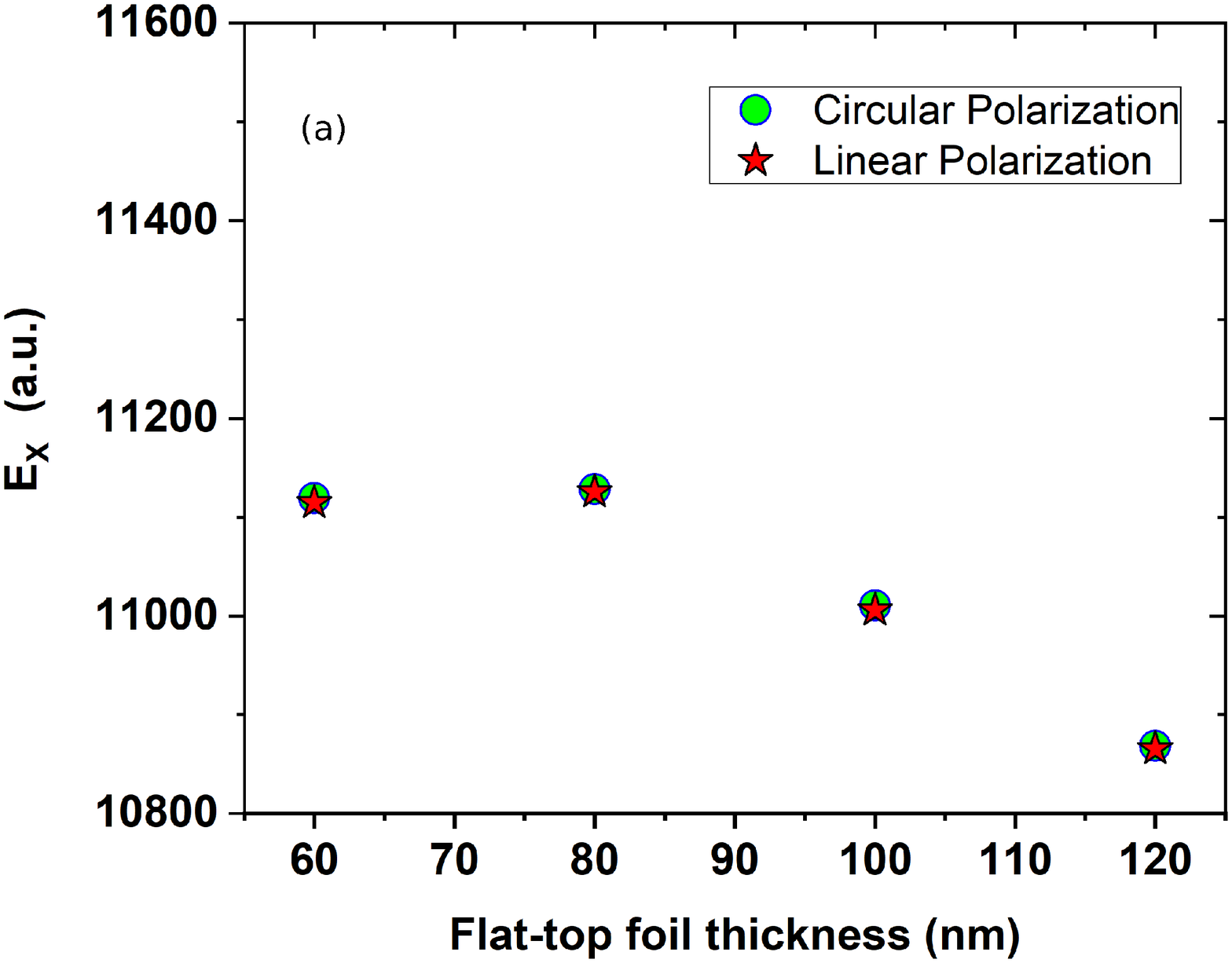} 
\hspace{5mm}
 \includegraphics[width=7cm]{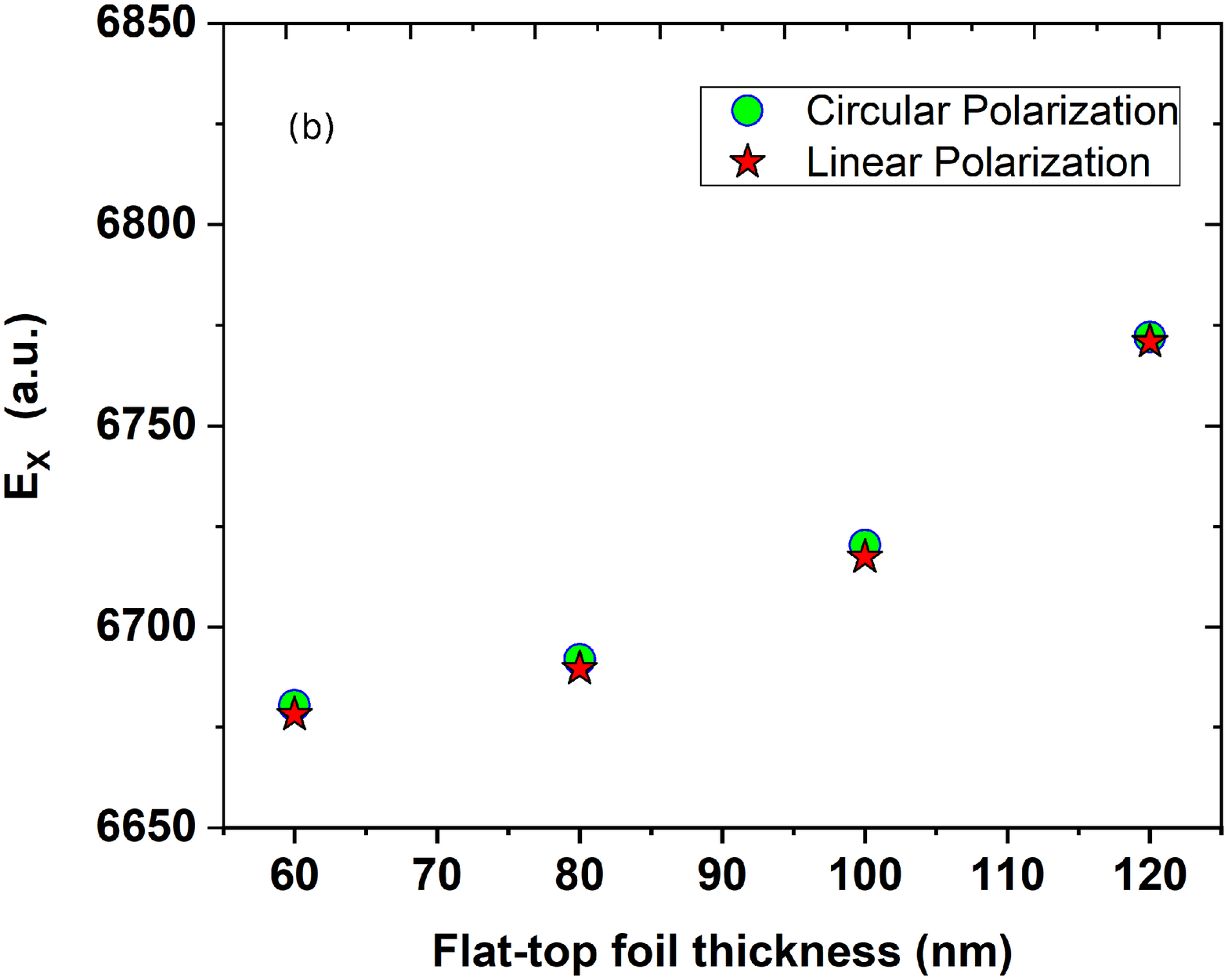} 
\caption{The behaviour of the electric field measured in the laser pulse $-$ nano-layer flat-top cone targets interaction point at (a) the moment t$_{0}$ and (b) the moment t$_{1}$.}
\label{fig:9}      
\end{figure*}

The geometry of the problem is depicted in Figure \ref{fig:8}. The FDTD numerical study presented here implies a Gaussian laser source with the same parameters as in the PIC simulations which propagates along the $x$-axis, to the fs pulses - target interaction point. The 5.6 $\mu$m source diameter is specified at a distance of 20 $\mu$m from the interaction point, along the propagation  $x$-axis, where the field is generated as initial condition. The material considered for the cones, foils and the nanospheres is plastic. This study together with the PIC numerical analysis previously described aims to provide a complex and useful approach for experiments proposed at ELI-NP.
The resulting field data in a certain observation point are plotted in Figures \ref{fig:9}(a) and \ref{fig:9}(b). The $x$-component of the electric field, E$_{x}$ is calculated in a specific observation plane which corresponds to the ultra-high intensity laser pulse $-$ target interaction point. The thickness of the nano-layered flat-top cone targets depends on the spheres and foil dimensions. Thus, the nano-foil thickness varies in the range of 60-120 nm with 20 nm step and nanospheres diameter varies in the range of 20-80 nm with the same step. Initially, the study has been elaborated using a nano-layer flat-top cone target with the nano-layer foil thickness in the range previously mentioned. The on-axis electromagnetic field was computed by the specific time monitor for both linear and circular laser pulse polarization, at a given moment of time, when the maximum value of the electric field is reached.
\begin{multicols}{2}
\end{multicols}
As depicted in Figures \ref{fig:9}(a) and \ref{fig:9}(b), the $x$-component of the electric field measured in the laser pulse $-$ nano-foil of the cone target interaction point reaches the highest value when the foil thickness is equal to 80 nm. This moment of time is noted with t$_{0}$. At the moment t$_{1}$, which occurs 27 fs after the moment t$_{0}$, the same temporal monitor indicates another maximum in the slope of the $x$-component of the electric field, with a different distribution of its highest registered values. On this moment of time, the highest value of the electric field corresponds to the nano-foil thickness equal to 120 nm (Figure \ref{fig:9}(b)). In both cases, the $x$-component of the electric field exhibits slightly higher values for circular polarization of the laser beam comparing to the linear one. The longitudinal component of the electric field is responsible for the ion acceleration. This can explain the similar behaviour of the maximum kinetic energies of protons and carbon C$^{6+}$ ions whith the one we obtained by performing PIC simulations.  
\begin{multicols}{2}
\begin{figure*}
\vspace{2mm}
\centering
 \includegraphics[width=7cm]{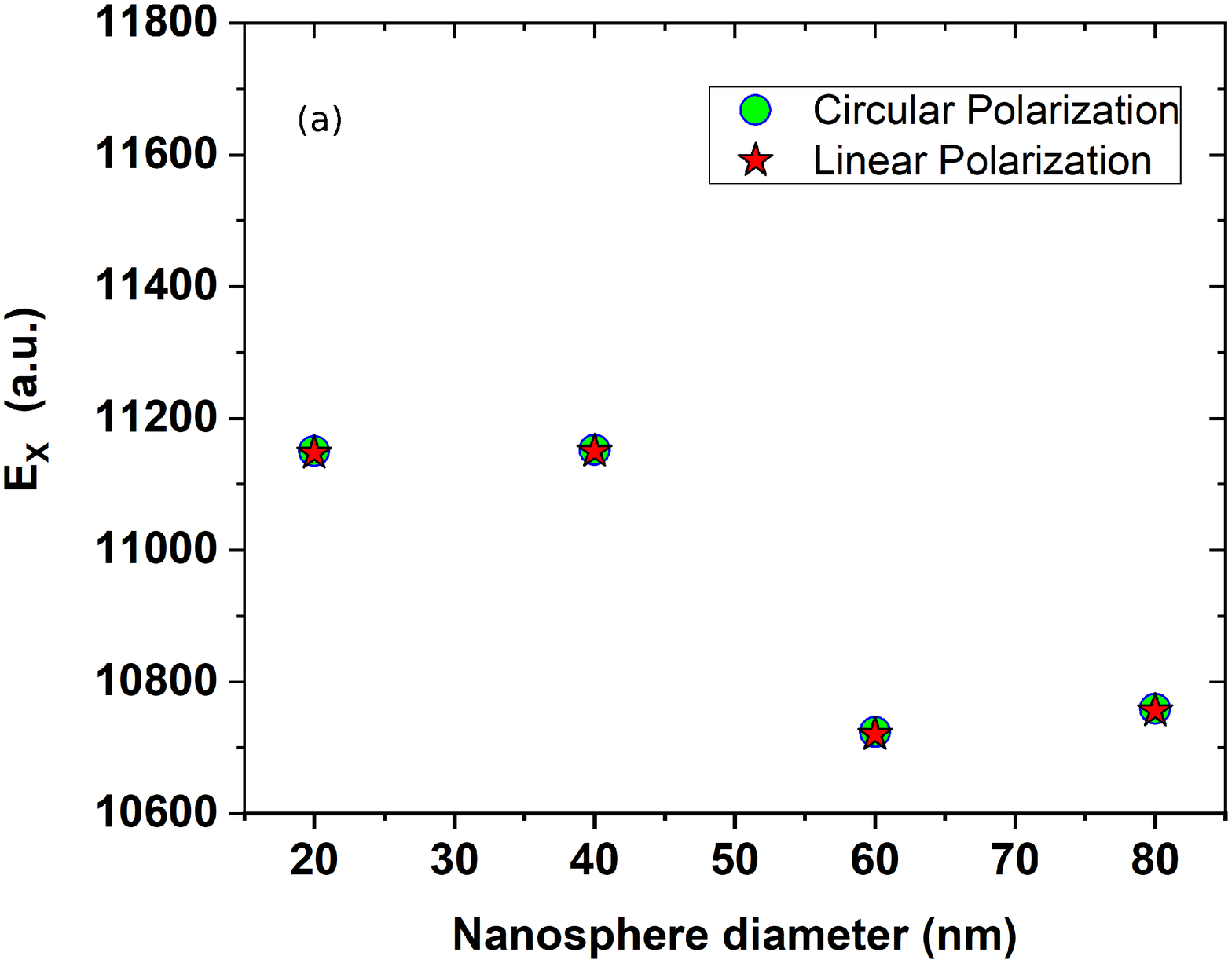} 
\hspace{5mm}
 \includegraphics[width=7cm]{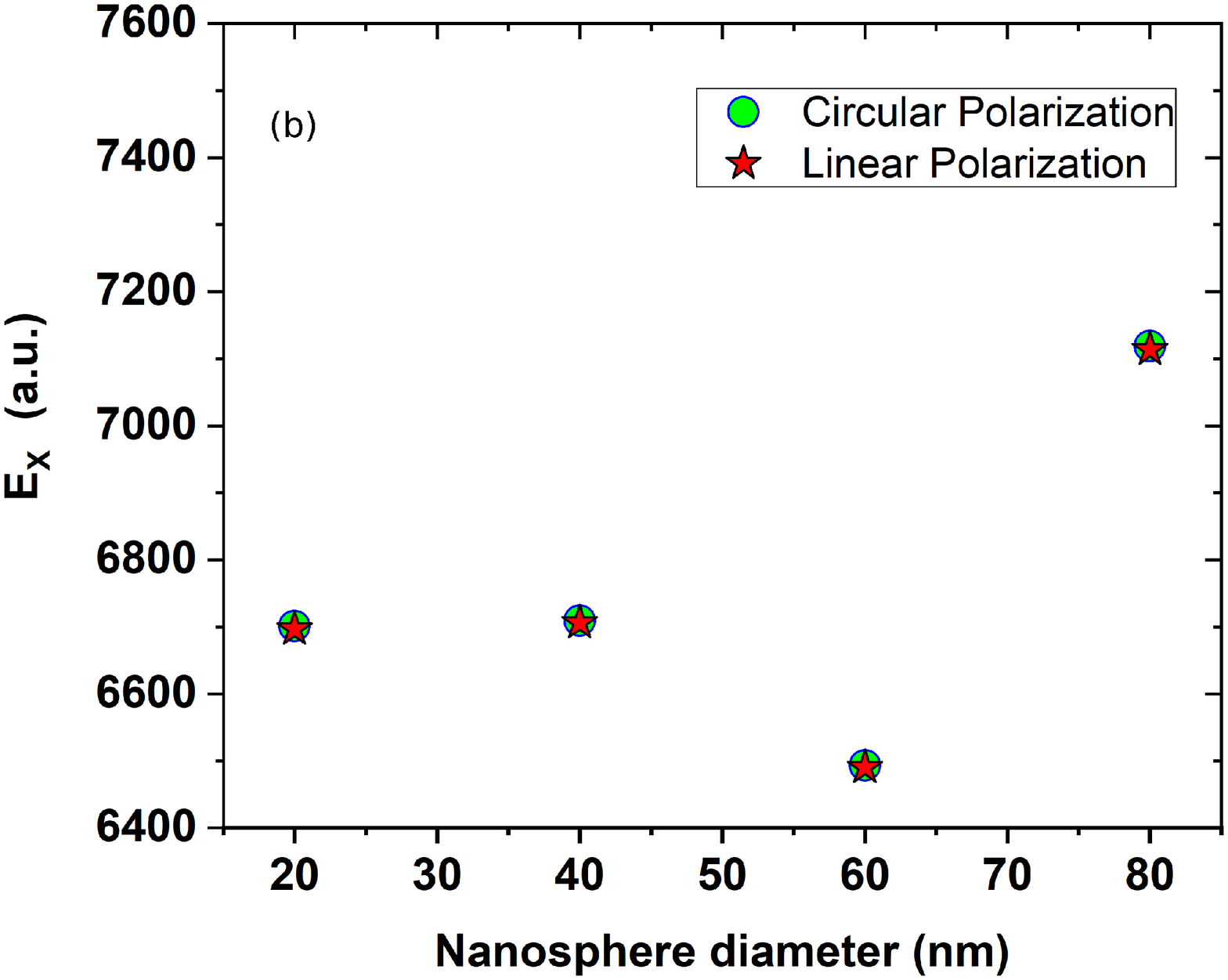} 
\caption{The behaviour of the electric field measured in the ultra-high intensity laser pulse $-$ double nano-layered flat-top cone target interaction point at (a) the moment t$_{0}$ and (b) the moment t$_{1}$.}
\label{fig:10}      
\end{figure*}
\end{multicols}
Comparing the case of the nano-layer flat-top cone target (Figure \ref{fig:9}) with the case of double nano-layered flat-top cone target (Figure \ref{fig:10}) for each nanospheres and nano-foil dimensions previously mentioned, we observe a similar electromagnetic field behaviour in terms of the maximum values obtained in the laser$-$matter interaction point. The highest registered value of the electric field at the moment t$_{0}$ corresponds to the nanosphere diameter equal to 40 nm while at the moment t$_{1}$, the electric field reaches the maximum value in the case of coated foil target with nanospheres having the diameter equal to 80 nm. Both these values are higher than the ones obtained for the case of nano-layer flat-top cone target. Again we have the same results as the ones we achieved by PIC simulations.

\section{Conclusions}
\label{concl}
We studied via numerical simulations, PIC simulations and FDTD method simulations, two new types of nanotargets which can be used for laser-ion acceleration. First target was a plastic flat-top cone target with the flat-top foil having the thickness of tens of nanometers. The second target was a plastic flat-top cone target with the flat-top foil composed by two nano-layers. One of the nano-layers consists of nanospheres with the same diameter of tens nanometers. The other one is a nano-foil with a thickness of tens nanometers, too. An ultra-high intensity circularly and linearly polarized laser pulse with the intensity of $2.16\times 10^{22}$ W/cm$^2$ interacts with these targets. For the ultra-high intensity laser pulse with circular polarization we obtained higher values of the maximum proton and carbon C$^{6+}$ ion energies than for the linear polarization. For all the target geometries and laser pulse polarizations there are monoenergetic beams of the localized protons with energies higher than 200 MeV. Therefore, these targets are appropriate for proton therapy of cancer. But only for the circularly polarized ultra-high intensity pulse and double nano-layers flat-top cone targets with the nanosphere diameter of 80 nm and the foil thickness of 40 nm the C$^{6+}$ ion beam is monoenergetic and the energy higher than 4 GeV. Hence, this target is good for hadron therapy. The protons and Carbon C$^{6+}$ ions are accelerated through the radiation pressure acceleration mechanism.             
We belive that the targets studied in our paper could be used in the future experiments of laser-ion acceleration at ELI-NP. 

\ack
This work has been financed by the national project PN III 5/5.1/ELI-RO No. 16-ELI/2017 (“SIMULATE”), under the financial support of Institute for Atomic Physics – IFA and by the NP Nucleu LAPLAS V 3N/2018, under the financial support of the Ministry of Research and Innovation.

\end{document}